\documentclass[aps,prl,preprint,groupedaddress]{revtex4}
\usepackage{graphicx}
\begin{document}
\title{Exchange Bias in Ferromagnetic/Compensated Antiferromagnetic Bilayers}

\author{D.S. Deng, X.F. Jin}
\affiliation{State Key Laboratory of Applied Surface Physics and
Department of Physics,Fudan University, Shanghai 200433, China}

\author{Ruibao Tao}
\affiliation{Chinese Center of Advanced Science and
Technology,(World Laboratory),P.O.Box 8730, Beijing 100080,
People's Republic of China \break Department of Physics, Fudan
University, Shanghai 200433, China}

\date{\today}

\begin{abstract}
By means of micromagnetic spin dynamics calculations, a
quantitative calculation is carried out to explore the mechanism
of exchange bias (EB) in ferromagnetic (FM)/compensated
antiferromagnetic (AFM) bilayers. The antiferromagnets with low
and high N\'{e}el temperatures have been both considered, and the
crossover from negative to positive EB is found only in the case
with low N\'{e}el temperature. We propose that the mechanism of EB
in FM/compensated AFM bilayers is due to the symmetry broken of
AFM that yields some net ferromagnetic components.
\end{abstract}

\pacs{75.70Cn, 75.30Gw}

\maketitle

Exchange anisotropy was first discovered in 1956 by Meiklejohn and
Bean\cite {first}, who found that the hysteresis loop of Co/CoO
after cooling in a magnetic field was no longer centered at zero
field (H=0) but was shifted along the field axis. The shifted
direction was found to be opposite to the applied magnetic field
(negative exchange bias $H_B<0$) and the magnitude of this shift
is known as exchange bias (EB). It was subsequently established
that this might be a general phenomenon for any ferromagnet
(FM)/antiferromagnet (AFM) systems cooling in an applied magnetic
field (cooling field) from above the N\'{e}el temperature(T$_N$)
of the AFM, with the FM Curie temperature(T$_C$) greater than
T$_N$. In recent years, since the phenomenon of exchange bias has
become the basis for an important application in information
storage technology \cite{Application}, tremendous efforts have
been made for exploring the mechanism \cite{Review},
\cite{Review2}.

Meiklejohn and Bean originally suggested that exchange bias was a
consequence of the presence of interfacial uncompensated AFM
spins. In view of this argument, a natural question to ask is
whether the exchange bias
also exists in a FM/ compensated AFM system. Surprisingly, in a compensated $%
Fe/FeF_2$ bilayer system, Nogues et al. observed not only the
usual negative exchange bias but also an unexpected positive
exchange bias ($H_B>0$) under large cooling fields
\cite{Positive}.

Several important theories have existed to study the exchange bias
in compensated AFM. Koon\cite{Koon} presented a microscopic
explanation of EB due to a irreversible AFM domain wall, and found
a perpendicular orientation between the FM/AFM axis directions,
namely spin-flop state. With consideration of magnetostatic
interactions in this spin-flop state, Schulthess et al.
\cite{Butler} obtained the opposite results, i.e., not EB but a
large uniaxial anisotropy, and attributed EB to the interfacial
defects. Unfortunately they did not further show the
EB-magnetostatic interactions phase diagram, in other words, EB
should gradually change with magnetostatic interactions.
Hong\cite{Hong} argued that interface spin configuration persisted
after cooling below T$_N$, and negative/positive bias respectively
corresponded to parallel/perpendicular easy axes of FM and AFM.
Kiwi et al.\cite{Kiwi} suggested a canted AFM spin configuration
frozen into a metastable state, and proposed the incomplete FM
domain wall model to explain positive exchange bias. However, the
former two theories were carried out with micromagnetic
calculations without consideration of the cooling field; the later
two theories pointed out the cooling field without micromagnetic
calculations, and were lack of much more detailed and sufficient
microscopic information. Up to now, exchange bias mechanism is
still controversial.

In this paper, based on the assumptions that an antiferromagnetic
interface coupling between FM/AFM is responsible for exchange bias
in FM/compensated AFM \cite{Positive}, \cite{Postive2}-
\cite{Coupling}, and that the biased hysteresis loop is basically
determined by the spin configurations in the underlying
antiferromagnetic layer after cooling \cite{Nature}-\cite
{asymmetry2}, we carry out micromagnetic calculations using spin
dynamics to explain the mechanism of EB in the FM/compensated AFM
systems. Physically the key point different from previous Koon's
micromagnetic calculations\cite{Koon}, addresses the cooling field
during cooling process. We succeed to reproduce both the negative
and positive EB effects. Qualitatively speaking, it is a
competition among (i) the cooling magnetic field (ii) the
interface coupling of FM/AFM, and (iii) the spin-spin interaction
and anisotropy of AFM, that eventually determines the spin
configurations in AFM during the cooling process. For an AFM with
weak spin-spin interaction (low T$_N$), the spin configuration of
AFM at low cooling magnetic field is dominated by the AF type
interface coupling of FM/AFM. Therefore, the initially compensated
AFM layers especially the interface AFM layer becomes weekly
uncompensated, resulting in a net ferromagnetic component opposite
to the cooling field (or the magnetization in FM). The hysteresis
loop is then measured at low temperature after removing the
cooling field, while the spin configuration in AFM is frozen.
Similar to the arguments given by Meiklejohn and Bean, it can
easily be deduced that the broken symmetry of AFM in this case
favors the negative exchange bias. However on the other hand, if
the cooling field $H_{CF}$ is large and becomes to dominate, then
a net ferromagnetic component along the cooling field is expected.
Because of the AFM type interface coupling of FM/AFM, it turns out
that the broken symmetry of AFM in this case favors the positive
exchange bias. On other hand, for AFM with strong spin-spin
interaction ( high T$_N$ ) only negative EB can be found in a
reasonable high $H_{CF}$. The quantitative results are given in
the following to reveal in details how these different terms
affect the broken symmetry of AFM layers and its correlation with
the exchange bias.

Our model Hamiltonian is

\begin{equation}
H=H_{A-A}+H_{F-F}+H_{A-F},  \label{Htotal}
\end{equation}
where $H_{A-A}$ is the part of AFM layers, $H_{F-F}$ and $H_{A-F}$
the FM and the interface coupling between AFM and FM layers. They
are
\begin{equation}
H_{A-A}={\sum_{<i,j>}}J_{A-A}{\bf S}_i\cdot {\bf
S}_j-D_A{\sum_i}\left( S_i^x\right) ^2-H_{CF}{\sum_i}\left(
g_Au_B\right) S_i^x,  \label{Hantifer}
\end{equation}

\begin{equation}
H_{F-F}=-{\sum_{<i,j>}}J_{F-F}{\bf S}_i\cdot {\bf S}_j-\ H_{CF}{\sum_i}%
\left( g_Fu_B\right) S_i^x,  \label{Hferro}
\end{equation}

\begin{equation}
H_{A-F}={\sum_{<a,f>}}J_{A-F}{\bf S}_a\cdot {\bf S}_f,
\label{Hinterface}
\end{equation}
where $g_A$, $g_F$, $u_B$, $D_A$ and $H_{CF}$ denote AFM Lande
factor, FM Lande factor, Bohr magneton, antiferromagnetic
anisotropy and cooling field in parallel with AFM anisotropy,
respectively. The exchange coupling among
spins is considered for nearest neighbor sites only. The subscripts $a$ and $%
f$ are associated with AFM and FM respectively. It is noticed that
the anisotropy of FM layer is neglected based on the fact that
most experiments did use the soft ferromagnets. The dipole-dipole
interactions in the system are not considered here, since it
affects only quantitatively rather than qualitatively on the
symmetry broken of AFM. As the previous models, we also assume
that $J_{A-F}$ $\sim J_{A-A}$ \cite{Koon}, \cite{Butler}. The
N\'{e}el temperature increases monotonically with $J_{A-A}$, thus
the interface coupling is stronger in the FM/AFM systems with
higher T$_N$ and vice versa.

Now we calculate the EB by the following spin dynamics approach\cite{Butler}%
, i.e.,the local effective field is determined from the gradient
of the energy, ${\bf H}_i^{{\bf eff}}=-\frac{\partial
H}{gu_B\partial {\bf S}_i}$, and $\left\{ {\bf S}_i\right\} $ is
required to satisfy the Laudau-Lifshitz equation of motion with
the Gilbert-Kelley form for the damping term:
$\frac \partial {\partial t}{\bf S}_i=gu_B{\bf S}_i\times \left( {\bf H}_i^{%
{\bf eff}}-\eta \frac \partial {\partial t}{\bf S}_i\right) $,
where $\eta $ denotes the damping parameter. This damping term is
phenomenological and is included to remove the energy from the
system and to ensure that the magnetic system is in a stable or
metastable equilibrium after sufficient iterating calculating
steps. A lattice with $50\times 50\times 2(FM)$ and 10 layers of
$AFM$ is used in our calculation. In the beginning, the
temperature $T$ of the system is set at $T_c>T>T_N$. Therefore,
the initial spin configuration in our spin dynamics calculation is
such that the spins are randomly arranged for AFM but are
ferromagnetic arranged for FM. A cooling field is then applied
along the $x$ direction that is also the easy axis of AFM.
Meanwhile it is known that the spins of FM will be easily aligned
according to the cooling field. After taking a long step of spin
dynamics calculation, a stable state of FM/AFM under the cooling
field is finally approached. Then the system is cooled down to the
low temperature, we switch off the cooling field and start to do
simulation of hysteresis loop of FM layers while the spin
configuration of AFM is fixed.

Fig.1 shows EB as a function of cooling field for a AFM/FM system
with low T$_N$ such as $FeF_2$ $(T_N\sim 78.4K)$ or $MnF_2$
$(T_N\sim 67.3K)$. In
doing this we set the parameters in Hamiltonian as $g_A=g_F=2.0,$ $%
J_{F-F}=10mev,$ $J_{A-A}=0.8mev,$ $J_{A-F}=J_{A-A}/2=0.4mev$ and
$D_A=0.4mev$ per site\cite{Review}, \cite{Koon}, \cite{Butler}. As
a natural output from the calculation, it is indeed observed in
this figure that the exchange bias $H_B$ changes sign from
negative to positive as the cooling field increases, and a
crossover field $\ H_{cross}$ is found at about $3.7T$. Dashed and
solid lines in the inset show the negative and positive loops at 2KOe and $%
7T$ respectively.

For systems with low T$_N$, i.e., weak spin-spin interaction
$J_{A-A}$ in AFM, the spin configuration of AFM at low cooling
magnetic field is dominated by AFM type interface coupling of
FM/AFM. In this case, the symmetry broken of compensated AFM
layers appears. Some net ferromagnetic component along the $-x$
axis is expected, which means that the broken symmetry of AFM in
this case favors the negative exchange bias. However for higher
cooling magnetic field $H_{CF}$, the cooling field becomes to
dominate the broken symmetry so that a net ferromagnetic component
along the positive $x$ direction is expected, i.e., the broken
symmetry of AFM in this case favors the positive exchange bias.

For a quantitative description of the FM components in AFM, We
define the ferromagnetic component in the n-th AFM layer

\begin{equation}
S^x(n)=\sum_iS_{na}^x(i)/N_n,  \label{Define}
\end{equation}
where $S_{na}^x(i)$ is the AFM spin at site $i$ of the n-th layer,
$N_n$ is the number of lattices in the n-th AFM layer, while the
first layer is defined as the interface layer of AFM. This
quantity $S^x(n)$ describes the degree of symmetry broken in each
layer of AFM. It is found from our calculation that the
ferromagnetic components are layer dependent. As expected, it will
be larger when the layer is near the interface, and become smaller
when the layer is far from the interface. In Fig.2 the
ferromagnetic component of the interface AFM layer $S^x(1)$ is
shown as a function of $H_{CF}$, using the same parameters as
obtaining Fig.1. Similar results as Fig.1 are found, that $S^x(1)$
is negative at the beginning when the cooling field is small, then
reaches to zero at a critical field $(\sim 37KOe)$ and finally
becomes positive as the field further increasing. Since the first
AFM layer should be the most important one in the interface
coupling, it is reasonable to see that the ferromagnetic component
of the interface AFM layer should be responsible for the EB
effect. The inset of Fig.2 gives the layer dependent ferromagnetic
components when the cooling field is fixed at 2KOe. The
oscillation is caused by the antiferromagnetic exchange
interaction between the layers of AFM.

Fig.3 shows the same relationship but for a AFM/FM system with higher $%
T_N $ such as FeMn $(\sim 500K)$. The parameters used here are $%
J_{F-F}=10mev,$ $J_{A-A}=5mev,$ $J_{A-F}=J_{A-A}=5mev$ and
$D_A=3mev$ per site \cite{Review}. In this case, it is found that
$H_B$ is always negative and changes a little when $\ H_{CF}$ is
ranged from 2T to 7T. This result also agrees qualitatively with
experiment\cite{Review}. Inset(a) of Fig.3 shows the magnetization
loop at 3T cooling field. In fact that $ J_{A-A}$ is large when
the AFM layer of the system has high $T_N,$ thus the interface
coupling $J_{A-F}$ also becomes large $\left( J_{A-F}\sim
J_{A-A}\right) $, then the AFM type interface coupling controls
the symmetry broken of AFM. In this case, if $H_{CF}$ is
reasonable high $(2T-7T)$ but not too high, $H_B$ is found to be
always negative. This can explain that
the positive EB was reported only in the FM/AFM thin films with low T$_N$%
\cite{Positive}, \cite{Postive2} - \cite{Coupling}.

One distinguished feature of Fig.3 is that: a tip EB is found at
cooling field $H_{tip}$ indicated by arrow, and this tip also is
clearly shown in Inset(b). As previous mention, the preceding
discussions are subjected to both cooling field and applied
magnetic field parallel with AFM easy axis. In fact, the
orthogonal FM/AFM spin configuration similar to Koon's conclusion
can also be recovered with zero or smaller cooling field for
stable spin configuration. With the increasing cooling field from
zero to $H_{tip}$, the FM spins will gradually rotate direction
from the perpendicular to parallel to the AFM easy axis during
cooling process, and the interface coupling contribution to the
negative bias will enlarge and nearly saturate at $H_{tip}$. On
the other hand, with the increasing cooling field above $H_{tip}$,
the cooling field contribution to the potentially positive bias
will raise, in other words, contribution to the negative bias will
lessen. Thus for cooling field parallel with AFM easy axis
situation, there exists a tip EB associated with cooling field
$H_{tip}$.

In summery, micromagnetic spin dynamics calculations are carried
out to explain the mechanism of EB in the FM/compensated AFM
system. Different from the previous micromagnetic calculations, we
address the key role of cooling field. Some important experiment
results, such as the cooling field dependent transition from
negative to positive EB in the AFM/FM layers with low T$_N$, can
be reproduced. It is proposed that the symmetry broken of AFM
plays a key role in explaining the exchange bias.

This work is supported by the National Natural Science Foundation
of China and Shanghai Research Center of Applied Physics. One of
the authors X.F. Jin would acknowledge the supports of the Cheung
Kong Scholars Program, Hong
Kong Qiushi Science Foundation, Y.D. Fok Education Foundation.

\begin{figure}[p]
\center{\includegraphics[width=16cm]{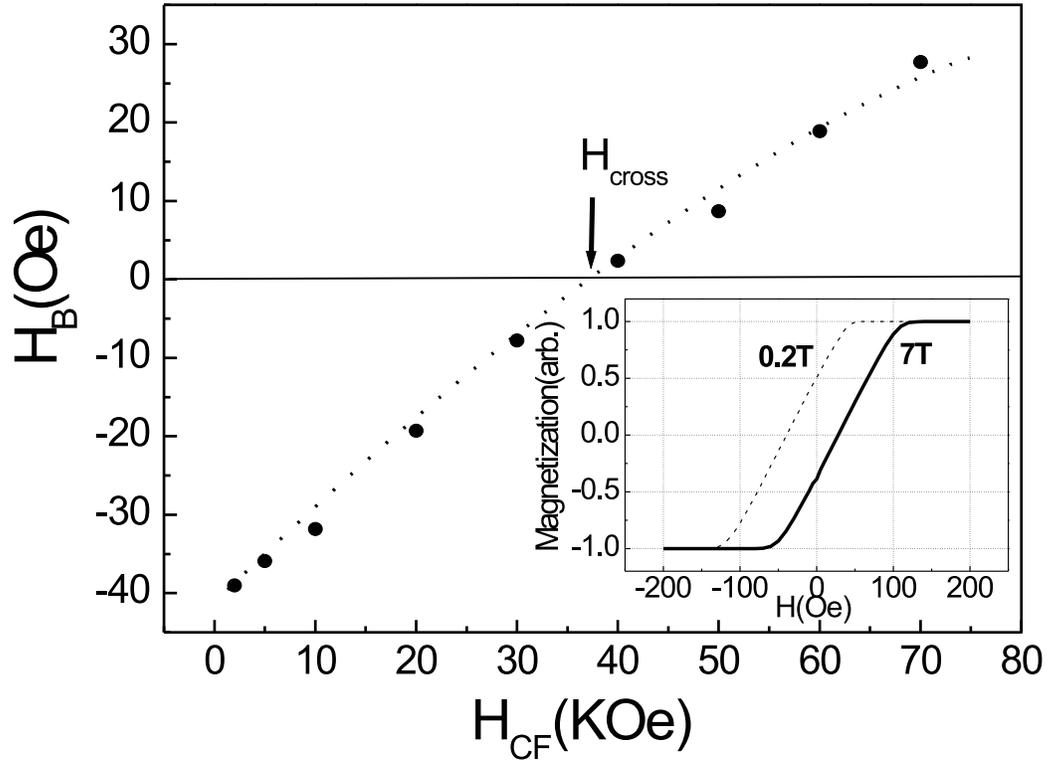}}
\caption{\label{fig:epsart}Exchange bias $H_B$ as a function of
cooling magnetic field $H_{CF}$ for FM/AFM with lower T$_N$
$(FeF_2).$ Dashed and solid lines in the inset show the negative
and positive magnetic loops at 2KOe and 7T cooling field,
respectively.}
\end{figure}

\begin{figure}
\includegraphics{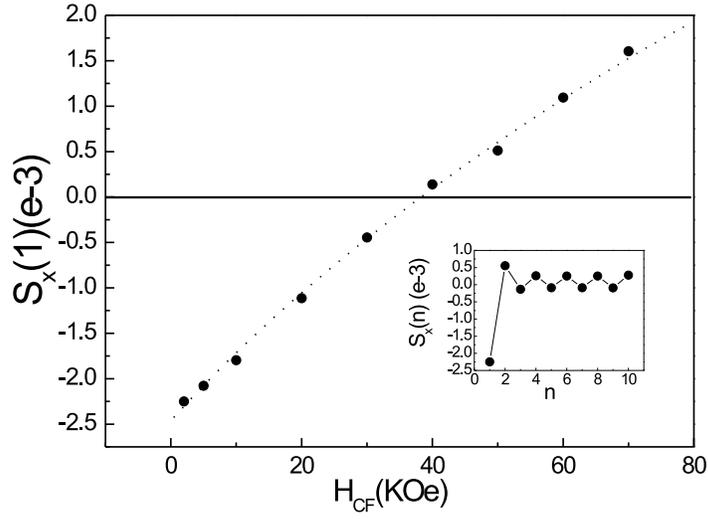}
\caption{\label{fig:epsart}The ferromagnetic component along the x
axis as a function of cooling field for the interface AFM layer.
Inset shows the layer dependent ferromagnetic components of AFM
layers at 2KOe.}
\end{figure}

\begin{figure}
\includegraphics{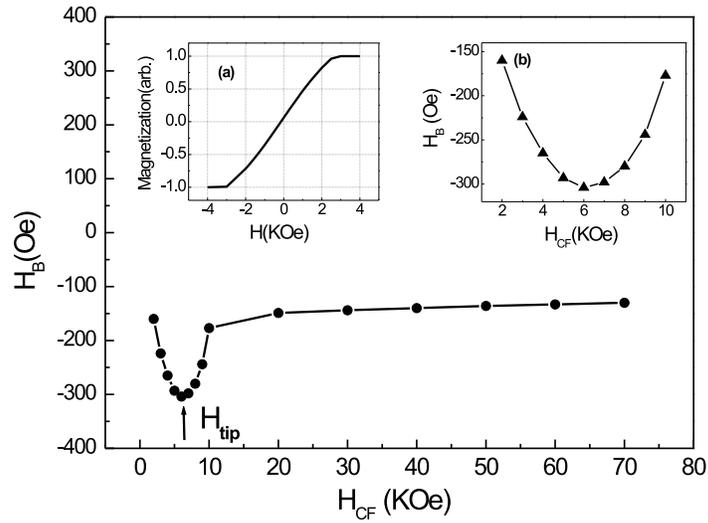}
\caption{\label{fig:epsart}The relationship between exchange bias
$H_B$ and cooling field $H_{CF}$ for FM/AFM with higher $T_N$
$(FeMn)$. Inset(a) presents a magnetic loop at 3T cooling field.
Inset (b) shows clearly the EB around cooling field $H_{tip}$
indicated by arrow in the figure.}
\end{figure}

\end{document}